\documentclass[12pt,english]{article}
\usepackage{graphicx}
\usepackage{setspace}
\usepackage{amsmath}
\usepackage{url}
\usepackage[authoryear]{natbib}
\usepackage[verbose,a4paper,tmargin=2.54cm,bmargin=2.54cm,lmargin=2.54cm,rmargin=2.54cm]{geometry}
\usepackage{lineno} 

\title{Seasonal species interactions minimize the impact of species turnover on the likelihood of community persistence}

\author{Serguei Saavedra$^{1}$\footnote{To whom correspondence should be addressed. E-mail: serguei.saavedra@usys.ethz.ch} \footnote{These authors contributed equally to this work} \footnote{Present address: Institute of Evolutionary Biology and Environmental Studies, University of Zurich, Winterthurerstrasse 190, CH-8057 Zurich, Switzerland and Environmental Systems Science, ETH Zurich, CH-8092 Zurich, Switzerland}, Rudolf P. Rohr$^{2}$\footnotemark[\value{footnote}] , Miguel A. Fortuna$^1$\footnote{Present address: Institute of Evolutionary Biology and Environmental Studies, University of Zurich, Winterthurerstrasse 190, CH-8057 Zurich, Switzerland}, \\Nuria Selva$^3$, and Jordi Bascompte$^1$\footnote{Present address: Institute of Evolutionary Biology and Environmental Studies, University of Zurich, Winterthurerstrasse 190, CH-8057 Zurich, Switzerland}
\\ 
\vspace{0.09 in}
\\$^1$Integrative Ecology Group \\ Estaci\'on Biol\'ogica de Do\~nana (EBD-CSIC) \\ Calle Am\'erico Vespucio s/n \\ E-41092 Sevilla, Spain\\
\\$^2$Department of Biology - Ecology and Evolution \\ University of Fribourg \\ Chemin du Mus\'{e}e 10 \\ CH-1700 Fribourg, Switzerland\\
\\$^3$Institute of Nature Conservation \\ Polish Academy of Sciences \\ Mickiewicza 33 \\ 31-120 Krakow, Poland}
\date{}

\begin{document}

\maketitle
\baselineskip=8.5 mm

\newpage
\begin{spacing}{1.9}
\raggedright


\section*{Abstract}

Many of the observed species interactions embedded in ecological communities are not permanent, but are characterized by temporal changes that are observed along with abiotic and biotic variations. While work has been done describing and quantifying these changes, little is known about their consequences for species coexistence. Here, we investigate the extent to which changes of species composition impacts the likelihood of persistence of the predator-prey community in the highly seasonal Bia\l owie\.{z}a Primeval Forest (NE Poland), and the extent to which seasonal changes of species interactions (predator diet) modulate the expected impact. This likelihood is estimated extending recent developments on the study of structural stability in ecological communities. We find that the observed species turnover strongly varies the likelihood of community persistence between summer and winter. Importantly, we demonstrate that the observed seasonal interaction changes minimize the variation in the likelihood of persistence associated with species turnover across the year. We find that these community dynamics can be explained as the coupling of individual species to their environment by minimizing both the variation in persistence conditions and the interaction changes between seasons. Our results provide a homeostatic explanation for seasonal species interactions, and suggest that monitoring the association of interactions changes with the level of variation in community dynamics can provide a good indicator of the response of species to environmental pressures.

\vspace{0.5 in}
{\sc Keywords: } Bia\l owie\.{z}a Primeval Forest, coexistence, ecological networks dynamics, food webs, homeostasis, predator-prey systems, structural stability\\

\newpage

\section*{Introduction}

Empirical evidence has revealed that networks of interspecific interactions are not permanent \citep{Margalef68,Ruiter}. Longitudinal studies have reported that ecological communities change not only their species composition across short and long timescales, but also the identity and strength of their interspecific interactions. Typically, these changes are observed along with seasonal environmental variation or an environmental gradient \citep{Houlahana}. For instance, in different trophic communities sampled across multiple years and seasons, studies have detected significant changes in species composition, percentages of diet consumption, number of interspecific interactions, and predator-prey ratios \citep{Baird,Schoenly,Tavares,Thompson,Hart}. Host-parasitoid and host-parasite communities have also been found to vary in both their interspecific interactions and species functional roles, to which can be attributed to temporal changes in species body size \citep{Laliberte,Pilosof}. Similarly, it has been well documented that the timing and length of species phenophase can lead to short and long-term interaction changes in mutualistic communities \citep{Petanidou,Olesen08,Alarcon,Carnicer,Diaz,Olesen11,Burkle}. 

\vspace{0.25 in}

While much work has been done looking at the description and quantification of interaction changes in ecological communities (see \cite{Poisot} for an extensive review of the literature), thus far, there is little agreement about both the driving mechanisms and the consequences of interaction changes for species coexistence \citep{Vizentin,Carstensen,Olito,Trojelsgaard}. Yet, understanding these effects is of paramount importance in order to face future community-wide risk scenarios of extinctions\citep{Tylianakis,Saavedra}.\\

\vspace{0.25 in}

Focusing on ecological communities subject to environmental variations, ecological theory suggests that the impact of seasonal changes can make a community to switch between different levels of maturity or development (e.g., seasonal differences in species composition) \citep{Margalef}. In particular, recurrent or periodic changes can preclude ecological communities from moving in a one-directional line of succession, and as a result, these communities can be typically found in an intermediate point in their developmental sequence. Estuaries, intertidal zones, and fresh water marshes are good examples of such communities, where species life histories are intimately coupled to the environmental periodicity \citep{Odum}. Because of this expected coupling, seasonal changes of species interactions are also expected to keep the community under a homeostatic state, i.e., under small variations in the conditions compatible with species coexistence despite changes of species composition in the community \citep{Odum,Ernest}.\\

\vspace{0.25 in}

To test the above hypothesis and to investigate any association between seasonal changes of species interactions and species coexistence, we study the seasonal dynamics of the terrestrial vertebrate predator-prey community in the Bia\l owie\.{z}a Primeval Forest (NE Poland). Between summer and winter, this community shows an important species turnover and changes in species interactions (predator diet). Using a general predator-prey model \citep{Case,Logofet,Rossberg}, we study the derived likelihood of community persistence as a function of both the network of species interactions and the changes in the community. This likelihood is estimated following a structural stability framework \citep{Thom,Alberch,Stone,Bastolla1,Bastolla,Rohr} and recent developments in the application of feasibility analysis to ecological \citep{Svirezhev,Logofet,Rohr} and non-ecological systems \citep{SaaRohr}. In particular, we investigate the extent to which the difference in species composition in Bia\l owie\.{z}a impacts the likelihood of persistence of the predator-prey community across the year, and the extent to which seasonal changes in species interactions modulate the expected impact.

\section*{Material and Methods}

\subsection*{Empirical data}

Bia\l owie\.{z}a represents the last old-growth temperate primeval forest in Europe, where seasonality is the organizing theme of the environment. The climate is continental with Atlantic influence, and two main seasons, cold and warm, are mainly distinguished \citep{Jedrzejewska}. Accordingly, predator diet has been often investigated separately in spring-summer (April-September) and in autumn-winter (October-March). For simplicity, we refer to them throughout the text as summer and winter, respectively. Winters can include periods of deep snow cover and extremely cold temperatures. The mean snow cover is $10$ cm, although in some winters it has reached up to $96$ cm and covered the ground from November until April. The temperature in January, the coldest month, averages $-4.8^\circ$C. June, July and August are the warmest months with mean daily temperatures of $17^\circ$C, and the highest insolation, on average 7 h/day compared to $48$ min/day in December \citep{Jedrzejewska}. Importantly, as expected, the strong seasonality in environmental conditions brings together an equally important species turnover between summer and winter \citep{Jedrzejewska}. Further details about the study area are provided in Appendix A.\\

\subsection*{Community characterization}

We characterized the community formed by predators (carnivores and raptors) and their prey in the Bia\l owie\.{z}a Primeval Forest (NE Poland) during the summer and winter seasons. Species presence and their interactions for the two seasons were compiled from 15 published studies covering two or more years mostly within the period 1985-1996 (see Appendix A). Data are available in Supplement A. In total, we observed 21 and 17 predators in summer and winter, respectively. All predators present in winter were also present in summer. We also observed 128 and 127 prey in summer and winter, respectively. From a total of 141 prey consumed across the two seasons, only 114 were consumed in both seasons. Species that are present across the entire year are called permanent species (e.g., black species in Figure 1), otherwise they are called seasonal species (e.g., colored species in Figure 1). This species turnover generated a summer ($N^S$) and a winter ($N^W$) predator-prey interaction network with seasonal species, and coupled by a subset of permanent species (see Figure 1). 

\vspace{0.25 in}

Importantly, interactions between permanent species were also changing between summer and winter. From a total of 435 interactions observed among permanent species, 303 were present in both seasons (e.g., black interactions in Figure 1), 62 in summer only (e.g., orange interactions in Figure 1), and 70 in winter only (e.g., blue interactions in Figure 1). This reveals that there is an important number of seasonal changes of interactions among permanent species that can be coupled to the environmental variations in Bia\l owie\.{z}a Primeval Forest.

\subsection*{Community dynamics}

To investigate the dynamics in the Bia\l owie\.{z}a predator-prey community in each season, we followed a general consumer-resource framework \citep{Levins,MacArthur,Case}. Traditionally, this framework has been used to develop reasonable explorations of predator-prey systems, while allowing the behavior of these systems to be analytically tractable and depend to a lesser extent on unknown parameters \citep{Case,Svirezhev,Logofet,Rossberg}. We described the dynamics of our predator-prey (consumer-resource) system by a general Lotka-Volterra model given by the following set of ordinary differential equations:
\begin{subequations} \label{equ:ode:system}
	\begin{align}
		\frac{dC_i}{dt} & = C_i \left( -m_i + \epsilon_i \sum_k \gamma_{ki} R_k \right) \\
    	\frac{dR_k}{dt} & = R_k \left( \alpha_k - R_k - \sum_i \gamma_{ki} C_i \right),
    \end{align}
\end{subequations}
where $C_i$ denotes the biomass of predator (consumer) $i$, $R_k$ denotes the biomass of prey $k$ (resource), $m_i$ is the mortality rate of predator $i$, and $\alpha_k$ is the intrinsic growth rate of prey $k$. Following previous work, $\epsilon_i$ is the standard conversion efficiency of predator $i$ and is set to $\epsilon_i=0.1$ \citep{Rossberg}. 

\vspace{0.25 in}

Additionally, $\gamma_{ki}$ denotes the trophic interaction strength between prey $k$ and predator $i$. We did not include intraguild predation because this is not observed in the data. For the sake of generalization and according to observations \citep{Margalef68,Saavedra}, we assumed that trophic interaction strengths are parameterized by $\gamma_{ki} = \gamma_0 / d_k^{\delta}$ when prey $k$ is consumed by predator $i$, and $\gamma_{ki} = 0$ otherwise. The parameter $\gamma_0$ represents the overall level of trophic interaction strength in the community. The variable $d_k$ denotes the number of predators consuming prey $k$, and $\delta$ is a scalable resource-partition parameter that modulates the consumption strength of prey $k$ among its predators. For each season, the elements $\gamma_{ki}$ are derived from the summer ($N^S$) and the winter ($N^W$) predator-prey interaction network accordingly (see Figure 1). Because the resource-partition parameter assumes a symmetric partition of prey's biomass among their predators, we also explored how asymmetric partitions affect our results (see Appendix B).

\subsection*{Likelihood of community persistence}

To investigate the likelihood of persistence of a predator-prey community, we studied the range of parameter space in the dynamical system (Equ. 1) leading to positive and stable biomasses for all species. The larger the range of parameter space compatible with positive stable solutions ($C_i^*>0$ and $R_k^*>0$), the larger the likelihood that the observed community can persist 
\citep{Rohr,SaaRohr}.

\vspace{0.25 in}

Because we are interested in positive stable solutions, first we needed to find the conditions leading to stability in the given system when subject to perturbations in species biomass. Following \cite{Case}, the dynamical system of Equation (1) does not have alternative stable states under a very large range of assumptions on its parameter values. It has been shown that for such dynamical systems, we can construct a Lyapunov function, which is a mathematical sufficient condition to constrain the dynamical system to converge to a single globally stable equilibrium point $(\boldsymbol{C}^*,\boldsymbol{R}^*)$. This implies that the dynamical system will absorb any perturbation in biomass and the system will eventually return to a globally stable equilibrium point. Therefore, the only question that remains to be answered is whether this predator-prey system can converge into positive stable equilibrium points, i.e., an equilibrium point with $C_i^*>0$ and $R_k^*>0$.

\vspace{0.25 in}

The conditions for having positive globally stable equilibrium points, once the interaction strengths have been fixed ($\gamma_{ki}$ and $\epsilon_i$), are dictated only by both the mortality rates of predators ($m_i$) and the intrinsic growth rates of prey ($\alpha_k$). The set of vectors [$\vec{m}$, $\vec{\alpha}$] compatible with stable persistence are the ones that make the solution of the following system of linear equations strictly positive:
\begin{subequations} \label{equ:feasibility}
	\begin{align}
		  \vec{m} & =  \text{diag}(\vec{\epsilon}) \boldsymbol{\gamma}^{t} \vec{R} \\
    	  \vec{\alpha} & =  \vec{R} + \boldsymbol{\gamma} \vec{C},
    \end{align}
\end{subequations}
where $\boldsymbol{\gamma}$ is the matrix of trophic-interaction strengths. This set of vectors, called the feasibility domain \citep{Logofet,Rohr,SaaRohr}, is given by 
\begin{equation} \label{equ:feasibility_domain}
\begin{split}
  D_F = \{ m_i & = \epsilon_i \gamma_{1i} x_1 + \cdots + \epsilon_i \gamma_{S_Ri} x_{S_R} \text{ and } \\
  \alpha_i & = x_i + \gamma_{i1} y_1 + \cdots + \gamma_{iS_C} y_{S_C} | \text{ with } x_i > 0 \text{ and } y_i>0 \}.
\end{split}
\end{equation}

\vspace{0.25 in}

Importantly, this feasibility domain is never empty: it is always possible to choose values for $m_i$ and for $\alpha_k$ such that we obtain a positive solution ($C_i^*>0$ and $R_k^*>0$). For instance, we can set $m_i = \epsilon_i \sum_k \gamma_{ki}$ and $\alpha_k = R_k + \sum_i \gamma_{ki}$, and the corresponding positive stable point is given by $R_k^* = 1$ and $C_i^* = 1$. This example stresses the importance of not only looking at whether the system can reach a positive solution, but also at how large the feasibility domain is (how big the set of vectors leading to a positive stable solution is). The larger the feasibility domain, the larger the likelihood of stable persistence. 

\vspace{0.25 in}

Building on previous work looking at competition systems \citep{Svirezhev,Logofet}, the size of the feasibility domain for predator-prey systems given by Equation (1) can be estimated by the following formula (see Appendix C for further details):\\
\begin{equation} \label{equ:ss}
	\Omega(\boldsymbol{\gamma}) = \frac{|\det(\boldsymbol{A})|}{ \prod_j (\sum_i A_{ij})},
\end{equation}
where the matrix $\boldsymbol{A}$, with its elements $A_{ij}$, is a two-by-two block matrix, function of $\boldsymbol{\gamma}$, given by
\begin{equation} \label{equ:A}
	\boldsymbol{A} = \left[
    \begin{matrix}
    	\text{diag}(\vec{\epsilon}) \boldsymbol{\gamma}^{t} & 0 \\
        I & \boldsymbol{\gamma}
    \end{matrix}
    \right],
\end{equation}
where $I$ is the identity matrix. This formula can be interpreted as the probability that a vector [$\vec{m}$,$\vec{\alpha}$] sampled uniformly at random (under the only constraint of being positive and with a fixed sum) falls within the feasibility domain. Therefore, the measure $\Omega(\boldsymbol{\gamma})$ can be used as a quantification of the likelihood of community persistence in summer and winter by using the corresponding matrix of trophic-interaction strengths ($\boldsymbol{\gamma}$) for each season.

\vspace{0.25 in}

Importantly, each likelihood $\Omega(\boldsymbol{\gamma})$ is a function of the overall trophic-interaction strength ($\gamma_0$) present in each season. For instance, for a value of $\gamma_0 = 0$ (no trophic interactions) predators cannot survive and consequently the likelihood of community persistence is zero. We found that the relationship between interaction strength and the likelihood of persistence is characterized by a concave function, meaning that there is a value of interaction strength ($\gamma_t$) at which the likelihood of persistence is maximized (see Fig. D1). Because we do not have the empirical data to infer $\gamma_0$ , we decided to use $\gamma_0=\gamma_t$ in order to calculate the maximum likelihood of community persistence for each season. Note that in order to calculate $\gamma_0$, we would require to have data on the amount of each prey biomass consumed by each predator. Importantly, the value of $\gamma_0$ does not change the qualitative results of our study (see Fig. D1). All calculations are performed using Matlab software version 2014a. Code in Matlab and R software are provided in Supplement B and C, respectively.

\subsection*{Expected likelihood of community persistence}

The expected likelihood of community persistence in a given season is evaluated from randomly-generated interaction networks. For each season, the only difference between the observed and randomized networks is the identity (not the number) of interactions among permanent species. These interactions in the randomized networks are randomly sampled from a meta network. Following ecological studies showing that many species interactions are forbidden due, for example, to phenological or morphological differences between species \citep{Jordano,Vazquez05,Olesen10}, the meta network is simply the source of all possible interactions that can be established between species according to our data. See Fig. D2 for an illustrative example of the meta network and randomizations. For each randomized network, we calculated the corresponding likelihood of community persistence as explained in the previous subsection. Using these likelihoods, we generated a distribution of expected likelihood of community persistence in summer and winter accordingly.

\subsection*{Observed variation in likelihood of community persistence}

To calculate the variation in the likelihood of community persistence between the observed summer and winter networks, we used the log absolute difference defined by $\text{log}(\Delta(\Omega)) = |\log(\Omega(\boldsymbol{\gamma}^S)) - \log(\Omega(\boldsymbol{\gamma}^W))|$, where $\Omega(\boldsymbol{\gamma}^S)$ and $\Omega(\boldsymbol{\gamma}^W)$ are the likelihood of community persistence for summer and winter, respectively.

\subsection*{Expected variation in likelihood of community persistence}

To calculate the variation in the likelihood of community persistence  between a randomized summer and winter networks, we used pairs of randomized summer and winter networks that share a given number of interactions among permanent species. The observed summer and winter networks share 303 species interactions. Thus, we generated randomized summer and winter networks as in the previous subsection, we selected pairs of randomized summer and winter networks that share 303 interactions, we  calculated their corresponding likelihood of community persistence as in the previous subsection, and we computed the variation in their likelihood as explained also in the previous subsection. Using these variations, we generated a distribution of expected variation in likelihood of community persistence. See Fig. D3 for an illustrative example of these randomizations.

\section*{Results}

\subsection*{Effect of species turnover on community persistence}

We find that the observed seasonal species turnover (i.e., differences in species composition) makes any potential combination of predator-prey interactions in the community to have a lower likelihood of persistence $\Omega(\boldsymbol{\gamma})$ in summer than in winter. Figure 2 shows that the observed predator-prey community has a lower likelihood of persistence in summer (orange line) than in winter (blue line). Figure 2 also reveals that the distribution of expected likelihood values generated by 50 thousand randomized winter interaction networks (right histogram) is always higher than the distribution of expected likelihood values generated by 50 thousand randomized summer interaction networks (left histogram). In other words, there are no such potential seasonal changes of species interactions that can compensate for the seasonal species turnover and keep the likelihood of persistence exactly invariant across the year. This shows that the likelihood of community persistence is strongly linked to the species composition present in the community. 

\vspace{0.25 in}

Additionally, these findings show that in summer the observed likelihood of persistence is large relative to the expected values within that season, while in winter, the likelihood is relatively small. We can interpret this result as a possible expression of an intermediate point in the developmental sequence of this community. As the species composition strongly fluctuates between summer and winter, the community may be pushed to maintain a closer connection between the resulting seasonal sub-communities.

\subsection*{Effect of seasonal interaction changes on community persistence} 

To investigate whether the observed seasonal species interactions modulate the impact of species turnover on the likelihood of community persistence, we measure how the observed variation in the likelihood of community persistence between summer and winter compares to the expected variations. We compare the observed variation to the situation where the number of randomly-generated permanent interactions is equal to the number of observed permanent interactions, and to the situation where there would be no interaction changes whatsoever. Because we do not know the direction of change in the data, i.e, from summer to winter or vice versa, we study both possibilities to eliminate interaction changes. The elimination of summer-to-winter change is generated by replacing all the interactions among permanent species in the winter network with the ones observed during summer. The elimination of winter-to-summer change is generated in the opposite way. The expected variation in the likelihood of persistence for the no change situation is then calculated between the summer (winter) network and the non-changed winter (summer) network.

\vspace{0.25 in}

Figure 3 shows that the observed variation in the likelihood of community persistence $\text{log}(\Delta(\Omega))$ (black/left line) is smaller than $99\%$ of 100 thousand pairs of randomly-generated expected variations (histogram). Similarly, Figure 3 shows that the observed variation is smaller than the expected variations generated from no interaction changes whatsoever (colored lines). These results reveal that the observed seasonal changes of species interaction minimize the variation in the likelihood of community persistence associated with species turnover across seasons.

\subsection*{Coupling of species to seasonal variations} 

To explain the community dynamics above, we analyze the extent to which individual species can be coupled to their seasonal environmental variations. Theory purports that mutual information between current and future states can reduce unnecessary changes that are energetically costly for species \citep{Margalef,Odum}. In this context, the fewer the changes in both the persistence conditions and the number of interaction changes, the higher the mutual information between seasons, and in turn, the higher the coupling between individual species and their environment.\\

\vspace{0.25 in}

To explore the above hypothesis, we quantify the extent to which the variation in the likelihood of of community persistence changes as function of the number of permanent interactions. Specifically, we compare the differences in the expected variations (e.g., see histograms of Fig. 3) when changing the number of randomly-generated permanent interactions between seasons. In the observed predator-prey community, 365 is the maximum possible number of permanent interactions between seasons, which is given by the seasonal network with the fewest number of interactions among permanent species (summer network). On the other hand, 70 is the minimum possible number of permanent interactions, which is given by the difference between the total number of interactions among permanent species (in both summer and winter) and the number of interactions among permanent species in summer. Recall that the observed number of permanent interactions is 303. Therefore, we vary the number of randomly-generated permanent interactions between 70 and 365.

\vspace{0.25 in}

We hypothesize that if lower expected variations would result from a larger number of permanent interactions than the observed number of permanent interactions, it would reveal unnecessary interaction changes in the Bia\l owie\.{z}a community. If much higher expected variations would result from a lower number of permanent interactions than the observed number, it would reveal a sub-optimized variation. Otherwise, it would reveal that species have indeed a strong coupling with the seasonality of their environment.

\vspace{0.25 in}

We find that all the distributions of expected variations, as function of the number of randomly-generated shared interactions, can be well approximated by a Gaussian distribution, and are characterized by the same mean (within $0.1\%$ of variation) but not by the same variance. This implies that the variance can be used as an indicator of how low or high the expected variation can move as function of the number of permanent interactions. 

\vspace{0.25 in}

Figure 4 shows that the larger the number of permanent interactions (viz., the smaller the number of interaction changes), the smaller the variance from the mean, and therefore, the larger the expected variations between seasons. In fact, as soon as we increase the number of permanent interactions from the observed value (dashed line in Figure 4), the variance exponentially drops. In contrast, decreasing the number of permanent interactions from the observed value can only marginally increase this variance. This reveals that the observed number of interaction changes sets the balance between reaching a low variation in the likelihood of community persistence and preserving a large number of permanent interactions between seasons.

\section*{Discussion} 

Ecological communities subject to abiotic and biotic variations have been typically characterized by temporal changes of interspecific interactions. Unfortunately, many of the consequences of these interaction changes are still poorly understood. To shed new light into these factors, we have studied the seasonal dynamics of the predator-prey community in the highly seasonal Bia\l owie\.{z}a Primeval Forest. Between summer and winter, this community shows an important species turnover and changes of predator diet. We have found that the observed species turnover generates a difference in the likelihood of community persistence across seasons regardless of any potential change of predator-prey interactions. Importantly, we have shown that the observed interaction changes minimize the variation in the likelihood of persistence across the year. These results support ecological theory suggesting that seasonal species interactions play a key role in maintaining a homeostatic state or a relatively low level of dynamical variation on ecological communities despite changes in species composition \citep{Margalef68,Odum,Ernest}.

\vspace{0.25 in}

Additionally, ecological theory suggests that simple rules of energetics and information can be governing the dynamics of ecological communities \citep{Margalef,Odum,Ruiter}. Here, we have shown that seasonal changes of species interactions are coupled to the environment by minimizing both the variation in persistence conditions and unnecessary dietary changes that can be energetically costly for individual species. Therefore, the observed community dynamics should not be understood as group selection, where the behavior of individual species would be expected to follow a common goal for all species. Under a group-selection framework, interaction changes would be an explicit mechanism of the entire community to achieve the goal of maintaining a low variation in the likelihood of persistence. In contrast, the intimate coupling between interaction changes and environmental seasonality can be simply the result of a long-term adaptation process on each of the individual species coping with seasonal changes and reducing unnecessary energetic costs.

\vspace{0.25 in}

By using a well-defined and parsimonious dynamical model, our findings represent a clear example of how seasonal changes of species interactions can have regulating effects on community persistence. This suggests that the adaptation of biological species to changing environmental conditions partially depends on their capacity to adjust their interspecific interactions. If interaction changes are slower or faster than the effects of environmental change, ecological communities may exhibit stronger fluctuations. While more detailed dynamical models can be incorporated, we advocate for the range of conditions leading to the stable coexistence of species as a potential quantitative measure of the likelihood of community persistence. We believe that monitoring the association of species interactions changes with the level of dynamical variation on ecological communities can provide a good indicator of the response of species to environmental pressures.

\vspace{0.25 in}

Finally, our results raise a number of interesting questions about the extent to which changes of species interactions along seasonal variations or environmental gradients should generate different consequences on ecosystem functioning. A potential hypothesis would be that relatively undisturbed ecological communities subject to seasonal or periodic environmental changes should exhibit relatively low variation in the likelihood of persistence across time. In contrast, under anthropogenically-generated changes or changes over an environmental gradient, ecological communities should exhibit relatively high variation in the likelihood of persistence. Such a hypothesis would be congruent with simulated effects of directional changes on ecological communities \citep{Saavedra}, the expected effects of global environmental change \citep{Tylianakis}, and requires further exploration.

\newpage

{\bf ACKNOWLEDGMENTS} 
Funding was provided by a postdoctoral fellowship, JAE-Doc, from the Program ``Junta para la Ampliaci\'{o}n de Estudios" co-funded by the Fondo Social Europeo (MAF), the European Research Council through an Advanced Grant (JB), and partly by the National Science Centre, project 2013/08/M/NZ9/00469, and the National Centre for Research and Development in Poland, Norway grants, POL-NOR/198352/85/2013 (NS). The funders had no role in study design, data collection and analysis, decision to publish, or preparation of the manuscript.
\\



\pagebreak

\bibliography{bibliography}{}
\bibliographystyle{ecology}

\pagebreak

\section*{Figure Captions}

\textbf{Figure 1.} Illustration of the Bia\l owie\.{z}a Forest predator-prey community in summer and winter. The top and bottom figures correspond respectively to a subsample of the summer and winter predator-prey interaction networks (see Appendix A for a matrix representation of the complete interaction networks). In each interaction network, predators are at the top and prey at the bottom. Species in black and color correspond, respectively, to permanent and seasonal species. Black and colored lines correspond to interactions among permanent species that are present the entire year (permanent interactions) and in one season only, respectively. Dashed lines represent interactions either between permanent and seasonal species or among seasonal species only. Permanent predators: wolf ({\em Canis lupus}), lynx ({\em Lynx lynx}), red fox ({\em Vulpes vulpes}), raccoon dog ({\em Nyctereutes procyonoides}), otter ({\em Lutra lutra}), polecat ({\em Mustela putorius}), northern goshawk ({\em Accipiter gentilis}). Summer predators: Eurasian badger ({\em Meles meles}) and lesser-spotted eagle ({\em Aquila pomarina}). Permanent prey: red deer ({\em Cervus elaphus}), wild boar ({\em Sus scrofa}), hare ({\em Lepus europaeus}), squirrel ({\em Sciurus vulgaris}), mouse and vole (Muridae, Arvicolidae), shrew (Soricidae), thrush ({\em Turdus} sp.), resident small passerine bird (Passeriformes), fish (Cyprinidae), and amphibian (Amphibia). Summer prey: migratory small passerine bird (Passeriformes), reptile (Reptilia), and insect (Coleoptera). Winter prey:  European bison ({\em Bison bonasus}).

\pagebreak

\textbf{Figure 2.} Species turnover impact the likelihood of persistence across seasons. The orange line (left) and blue line (right) show the observed persistence likelihood $\Omega(\boldsymbol{\gamma})$ in summer and winter, respectively. The left and right histograms correspond, respectively, to the expected persistence likelihood in summer and winter. Expected persistence likelihood values are generated from potential randomized interaction networks. The figure is generated using a resource-partition parameter $\delta=1$. Other values generate qualitatively similar results (see Figs. D4-D5).

\vspace{0.5 in}

\textbf{Figure 3.} Seasonal changes of species interactions modulate the variation in the likelihood of community persistence. The black line (left) corresponds to the observed variation in the likelihood of persistence between summer and winter (derived from Figure 2). The histogram corresponds to the population of expected variations in the likelihood of persistence from all potential pairs of randomized interaction networks in summer and winter that share 303 predator-prey interactions. The orange (right) and blue (middle) lines correspond, respectively, to the variation in the likelihood of persistence that would be expected in the scenario without interaction changes from summer to winter and vice versa. The figure is generated using a resource-partition parameter $\delta=1$. Other values generate qualitatively similar results (see Figs. D6-D7).

\pagebreak

\textbf{Figure 4.} Community balance between number of permanent interactions and variation in the likelihood of persistence. The figure shows the variance in the expected variation in likelihood of community persistence as function of the number of permanent interactions used to generate the randomized summer and winter networks. The mean value of the expected variation is the same (within $0.1\%$ of variation) across all the different number of permanent interactions. The dashed line corresponds to the observed number (303) of permanent interactions. Note that the larger the number of permanent interactions (i.e., the smaller the number of interaction changes), the smaller the variance of the expected variation (i.e., the lower the chances of reaching a low variation). The figure is generated using a resource-partition parameter $\delta=1$. Other values generate qualitatively similar results (see Figs. D8-D9).


\clearpage

\setcounter{figure}{0}

\begin{figure}[ht]
\centerline{\includegraphics*[width= 1.1 \linewidth]{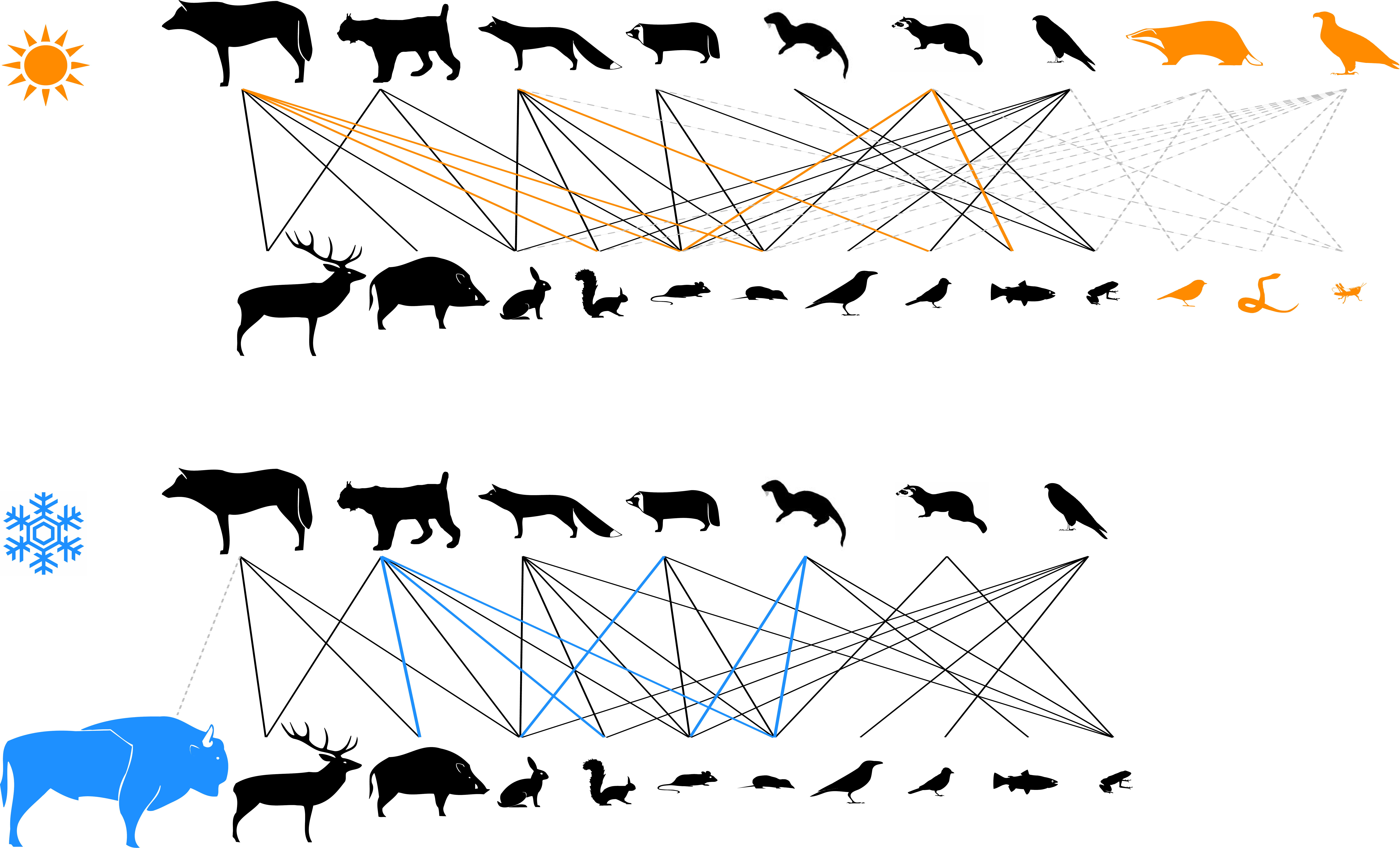}}
	\caption{}
\label{fig1}
\end{figure}

\pagebreak

\begin{figure}[ht]
\centerline{\includegraphics*[width= 1.0 \linewidth]{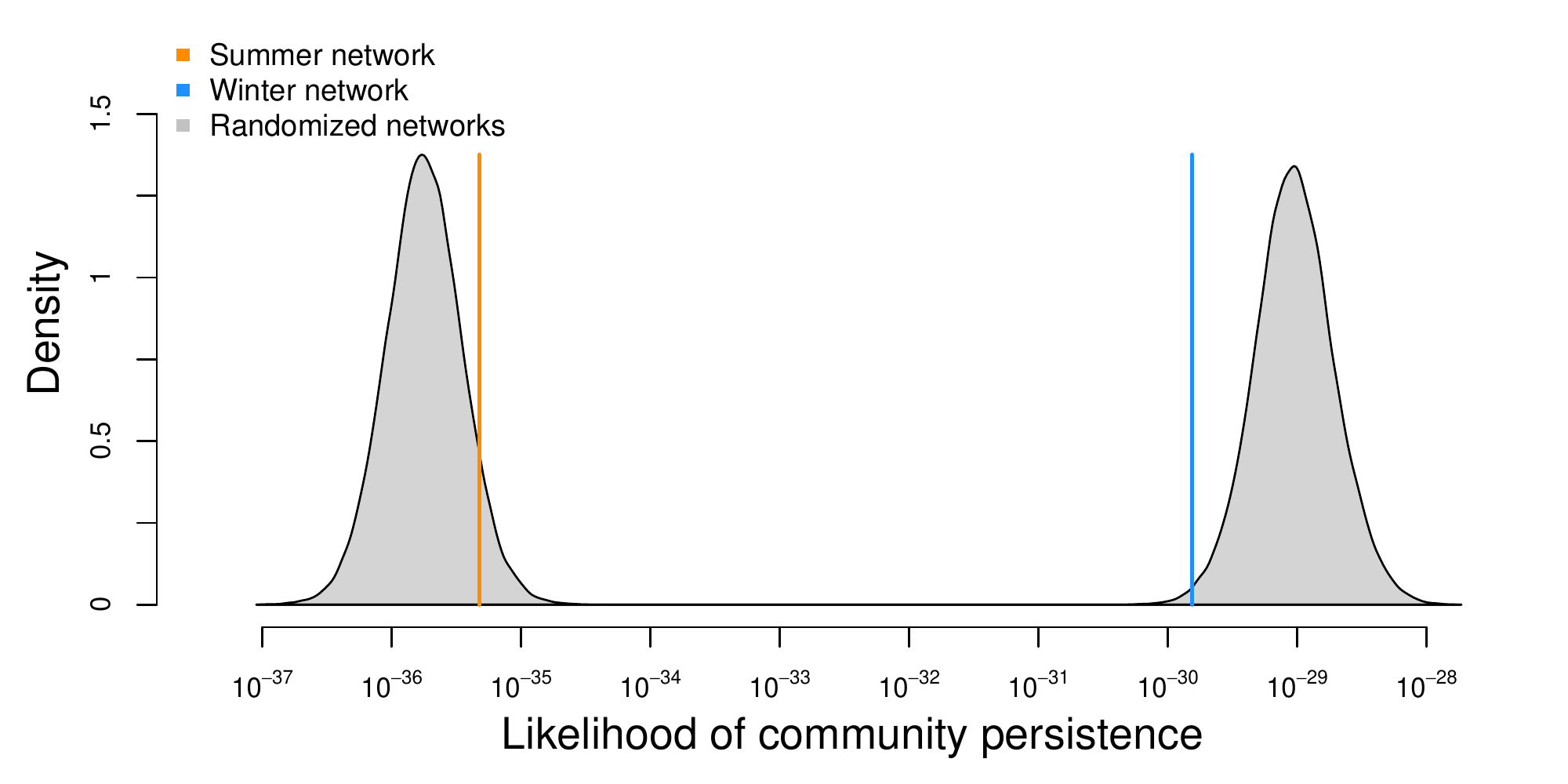}}
	\caption{}
\label{fig2}
\end{figure}

\pagebreak

\begin{figure}[ht]
\centerline{\includegraphics*[width= 0.8 \linewidth]{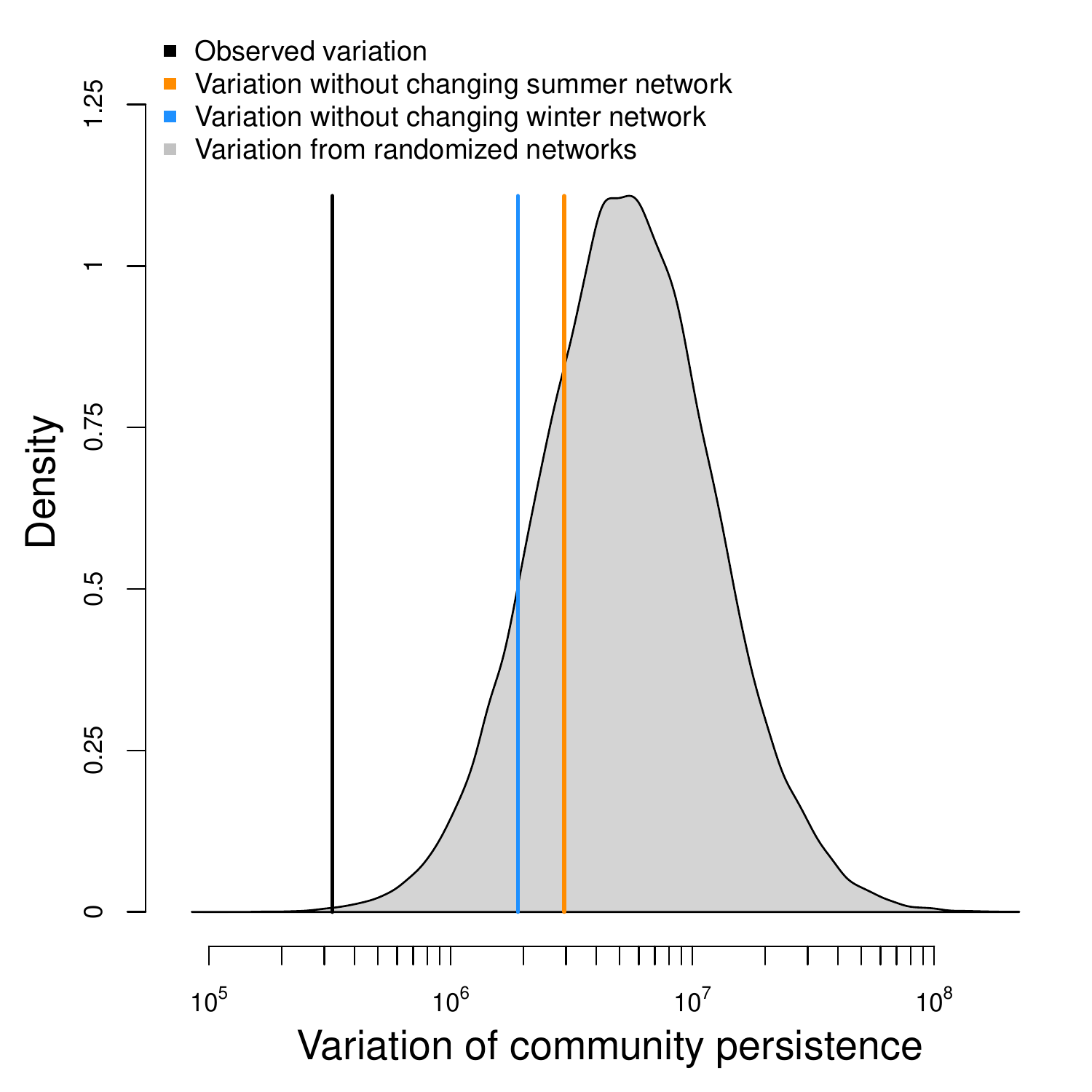}}
	\caption{}
\label{fig3}
\end{figure}

\pagebreak

\begin{figure}[ht]
\centerline{\includegraphics*[width= 0.8 \linewidth]{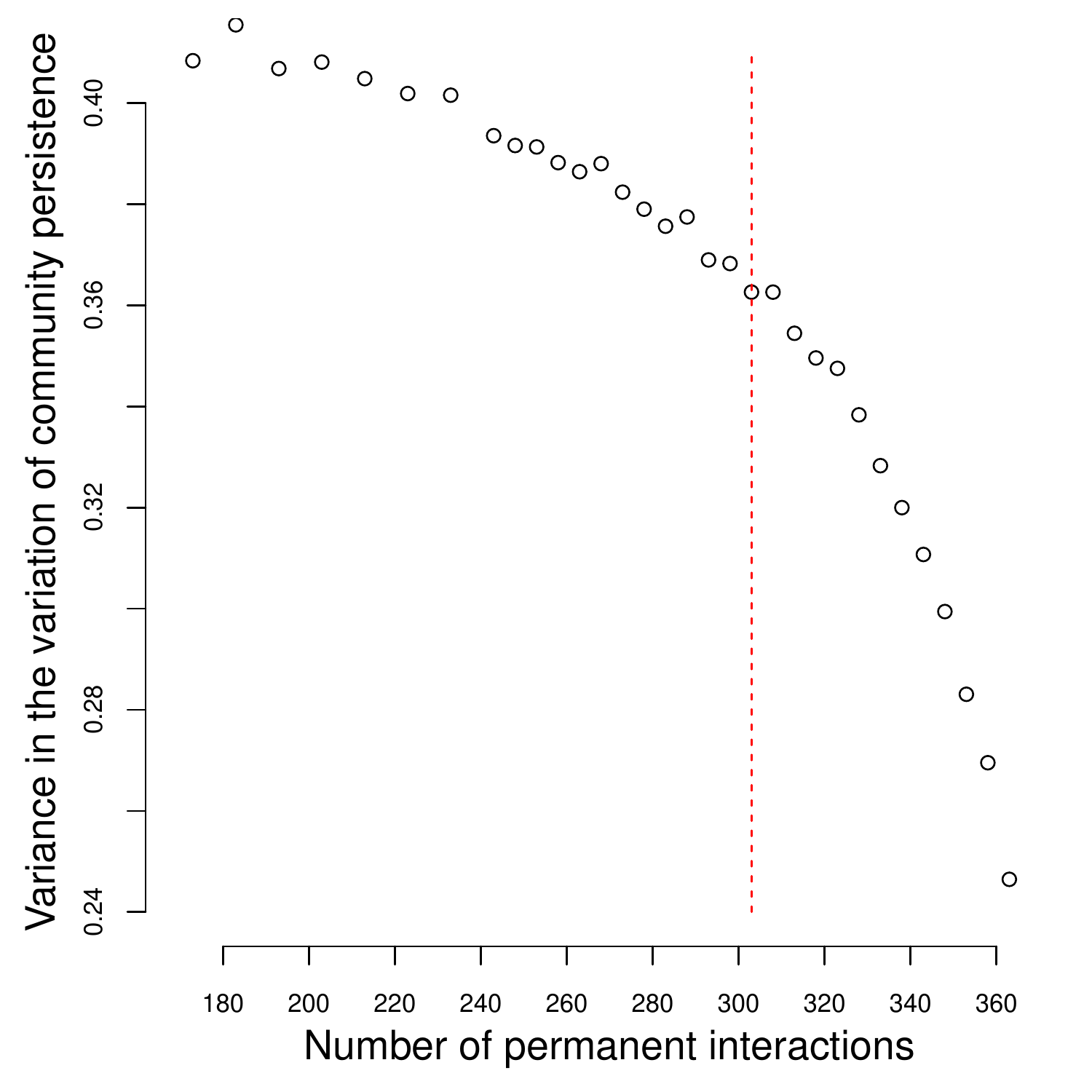}}
	\caption{}
\label{fig4}
\end{figure}

\end{spacing}
\end{document}